# Measuring nanoscale magnetic write head fields using a hybrid quantum register

I. Jakobi[1], P. Neumann[1], Y. Wang[1], D. Dasari[1], F. El Halak[2], M. Bashir[2], M. Markham[3], A. Edmonds[3], D. Twitchen[3] and J. Wrachtrup[1]

[1] *3. Physikalisches Institut, Universität Stuttgart and Institute for Integrated Quantum Science and Technology IQ$^{ST}$, Pfaffenwaldring 57, 70569 Stuttgart, Germany*

[2] *Seagate Technology, Londonderry BT48 0BF, UK*

[3] *Element Six Innovat, Didcot OX11 0QR, Oxon, England*

(Dated: February 8$^{th}$ 2016)

The generation and control of nanoscale magnetic fields are of fundamental interest in material science[1] and a wide range of applications[2]. Nanoscale magnetic resonance imaging[3,4] quantum spintronics[5,6] for example require single spin control with high precision and nanoscale spatial resolution using fast switchable magnetic fields with large gradients[7,8]. Yet, characterizing those fields on nanometer length scales at high band width with arbitrary orientation has not been possible so far. Here we demonstrate single electron and nuclear spin coherent control using the magnetic field of a hard disc drive write head. We use single electron spins for measuring fields with high spatial resolution and single nuclear spins for large band width measurements. We are able to derive field profiles from coherent spin Rabi oscillations close to GHz in fields with gradients of up to 10 mT/nm and measure all components of a static and dynamic magnetic field independent of its orientation. Our method paves the way for precision measurement of the magnetic fields of nanoscale write heads important for future miniaturization of the devices.

Precision control of magnetic fields on nanometer length scales is key to a variety of techniques most notably magnetic storage devices. Here, fields with a strength on the order of 1T and length scales of less than 100 nm are switched at GHz bandwidth. Such fields are also essential elements for precision coherent control of spins on nanoscale dimensions in quantum spintronics[9,10] or nanoscale sensing[11,12]. However characterising them remains an open challenge and a major obstacle for further miniaturisation. Single spins have the potential to measure such fields and field gradients precisely. Indeed, single spin probes like the nitrogen vacancy defect center in diamond have been used to measure static[13,14] and dynamic magnetic fields[15,16] with nanoscale resolution[17,18,19,20]. However, measuring large field gradients at high background field over a large bandwidth, essential for nanoscale spintronics, remains an open challenge. Here, we combine strong driving of single nuclear and electrons spins in large background fields with nanoscale field imaging by using techniques like Rabi imaging[21]. By combining the high band width of the write head field and coherent control of the nuclear spin we demonstrate a method to measure large magnetic fields at arbitrary orientation with single defect center spins. We believe that our method is a unique tool for further development of actively driven nanoscale magnetic field sources as they are found in e.g. write heads of magnetic storage devices[22].

The spin system we use to demonstrate fast driving and nanoscale field gradient imaging are single nitrogen vacancy (NV) defect center spins in diamond. The NV center is a ground state paramagnetic system with total electron spin S = 1. The three sublevels $m_S$ = 0, ±1 are split in an external magnetic field ($B_0$) by their Zeeman effect according to the spin Hamiltonian $H = hDS_z^2 + g\mu_B BS = H_z + H_\perp$ with $H_z = hDS_z^2 + g\mu_B B_z S_z$ and $H_\perp = g\mu_B(B_x S_x + B_y S_y)$. Here D marks the interaction of the two electron spins of the S = 1 state, g is the electronic g-factor and $\mu_B$ is Bohrs magneton. Precise measurement of the splitting of the three energy levels following from the Hamiltonian H provides information on the magnitude and direction of B. As the g factor is almost isotropic, all directional information is provided by D. It can be regarded as an effective magnetic (dipole) field with magnitude around 100 mT oriented along the ⟨111⟩ axis of the diamond crystal in which the electron spin is quantized. Using this anisotropy field, it is possible to reconstruct the full vector information of the magnetic field[13, 14, 23]. However, the spin anisotropy term not only generates an effective magnetic field but also influences the excited state dynamics of the defect and hence spin polarization in the ground state in such a way that optically detected spin measurements get challenging for off axis fields $B_{x,y}$ larger than 20 mT[24].

The basic physics of the effect have been quantitatively analysed[25] and used in previous imaging experiments[14]. In short, a transversal magnetic field causes spin state mixing as new states |i⟩ emerge from zero field states |j₀⟩ as $|i\rangle = \sum \alpha_{ij} |j_0\rangle$. Here, the coefficients $\alpha_{ij}$ are calculated from diagonalizing the Hamiltonian with the appropriate orientation and magnetic field strength $B_0$. The corresponding rates and the effective fluorescence intensity can be calculated from these new eigenfunctions by an averaging using the weights of the initial and the final states of the respective transition. In essence, any non-axial field $B_\perp$ results in a drop of the steady state fluorescence intensity with varying weight for different axial fields $B_z$. While generally there is an ambiguous dependence at higher fields ($|B| \gg 100$ mT) changes in the field strength $|B| = \sqrt{B_z^2 + B_\perp^2}$ have a negligible effect and the intensity mostly depends on the field angle $\theta = \arctan(B_\perp/B_z)$ (see Fig. 1d). In addition to the steady state the decay rates also influence the lifetime of the excited state and consequently the time-dependent behavior of the fluorescence after an excitation pulse[25].

To generate large, switchable magnetic field gradients we use write head structures fabricated to generate magnetization patterns in the magnetic recording medium of hard disk drives. The structures consist of a material with high permeability (write pole) surrounded by a coil to generate the magnetic field as shown in Fig. 1a,b. To shield off stray fields and constrain the field of the central write pole it is surrounded by a shield of identical material (return pole in Fig. 1b). The size of the write pole is around 100 nm. One return pole wraps around the write pole in three directions with a minimum gap of 20 nm. The second return pole is located in a 2 μm distance on the remaining side. Fig. 1c shows a finite element simulation of the magnetic field of the device in a plane parallel to the surface at 10 nm distance from the write pole. The simulation includes the return pole. As can be seen from the simulation the maximum field amplitude is on the order of 1 T. In the direction of the proximal return pole (top) it decays over a distance of 100 nm to almost zero. Field gradients along this direction are around 10 mT/nm. The band width of the device is around 3 GHz, i.e. large enough to not only generate a static magnetic field but also to directly generate electron and nuclear spin Rabi oscillations. In experiments, the head is directly placed on top of the diamond[26]. NV centers with a distance of around 10 nm below the surface were used for monitoring the field generated by the write pole (see Fig. 1b).

Figure 2 shows photoluminescence maps of a single NV scanned under the write head at an axial distance of around 50 nm[26]. Rather than continuously recording the steady state fluorescence we apply a train of excitation pulses and accumulate the signal during gates in the subsequent decays. This increases the contrast[27] and reduces short-lived background fluorescence from the write structure[26].

We recorded images for two different orientations of NV under the write head, i.e. for an NV in a diamond with a {100} and {111} surface orientation. In the first case, the NV axis is tilted 54.7° from the surface normal. NVs for the {111} surface are parallel (or antiparallel) to the surface normal. A characteristic feature of both measured photoluminescence maps are bright areas with a diameter of around 50 nm, which are also reproduced in the simulations[26]. In these spots, the orientation of the NV axis is within 20° parallel to the magnetic field of the head. Both, the shape of the bright spots and their position is different for the two different spin orientations and once again are in reasonable agreement with simulations. A further characterization of the field is made by taking line scans through the images and compare them with simulations of the magnetic field, both for {111} and {100} orientations. Assuming an offset field of 200 mT, changes in the field strength $|B|$ for the line scan can be neglected and intensity changes are only attributed to changes in the field angle $\theta$ relative to the NV axis[26] (see Fig. 2e). The line scans once again show that the two different NV orientation measure the field at different locations. They also demonstrate that for both cases the change in field orientation upon scanning the head over the defect is different. For the {100} orientation, a displacement of 10 nm changes the angle between NV and B by $\Delta\theta = 10°$ with a concomitant reduction of the photoluminescence by 22 %. In this case, changes in field orientation can be measured with a precision of a few degrees. The {111} orientation measures the field right under the write pole where the field has a lower gradient compared to its edges. As a result, the changes in the angle as well as the photoluminescence around the alignment position are smaller than for {100} reaching $\Delta\theta = 5°$ and 10 % within a 10 nm step. At the spots, optically detected magnetic resonance can be recorded to measure $|B|$ (see Fig. 2f). For both orientations, the field saturates above 200 mT. While {111} reaches a value of 207 mT the {100} oriented NV measures a B up to 218 mT. This is consistent with simulations for 60 and 40 nm axial distance between write pole and NV center, respectively.

To measure large off-axis fields, we use the broad bandwidth of the write head structure. We devised a scheme in which the spin is initialized and read out with the field switched off, $B_{x,y,z}=0$ (see Fig. 3a). The timing between initialization and application of B guarantees that the field is only applied when the defect is in its ground state. In the first part of the experiment, we use the write head field to generate Rabi oscillations of the defect electron spin. For this an AC magnetic field in resonance with the transition frequency of the defect is applied, i.e. at 2.87 GHz, after the spin has been initialized with a 3 µs laser pulse in its $m_s$=0 sublevel. The $B_x$ (or $B_y$) component of the field generate Rabi oscillations with frequency $\Omega = \gamma B_x/\hbar$ [26]. After the driving field is switched off, the spin state is read out by a laser pulse. Fig. 3a shows the result of Rabi frequency measurements for different driving fields and displacements of the write pole and the defect. Apparently, with this technique we are not restricted to a narrow range of field orientations but can measure the field at any location. Fig. 3a shows that the B-field amplitude is significantly reduced from its maximum value under the pole. The change is particularly pronounced along the direction in which the shield is closest to the write pole. The feature in the line scan has a width of $\Delta x = 292$ nm. With the microwave wavelength of $\lambda = 10.4$ cm this measurement signifies a resolution of $\Delta x/\lambda = 2.8 \cdot 10^{-6}$ in terms of near-field microscopy[28]. From the same measurement a magnetic gradient of $dB_x$ = 18 µT/nm can be derived. Even though we did not maximize the current amplitudes and the distance to the pole is around 50 nm this value is comparable to the largest gradients used so far for nanoscale magnetic resonance imaging[3, 7, 19]. The maximum Rabi frequency measured was 0.333 GHz which was not limited by the drive field but by the timing precision of our equipment.

To simultaneously measure $B_z$ we slightly modify our measurement sequences. To this end we use a Ramsey sequence and measure the resulting Larmor precessions of the NV spin that are induced by the axial component of the field $B_z$ by introducing a $\pi/2$ pulse after initialization. The spin state $|\psi\rangle = \frac{1}{\sqrt{2}}(|0\rangle + e^{i\phi}|1\rangle)$ after the $\pi/2$ pulse with $\phi = \frac{g\beta}{\hbar}\int_0^\tau B_z(t)dt$. After application of the field, a second $\pi/2$ pulse converts this phase into a measurable population which are read out by the final laser pulse. Fig. 3b shows the experimental result with a step-wise increase of $B_z$. With this method we are able to measure $B_z$ up to 18 mT for any component of $B_{x,y}$.

As a result, all three components of the magnetic field with arbitrary field orientation can be measured. However, with increasing field the signal frequencies quickly scale beyond the capabilities of our measurement hardware. In particular, for fields up to 200 mT timing resolution needs to be better than 100 ps. By choosing a single nuclear spin with lower gyromagnetic ratio for field measurement[29] one can keep spatial resolution while measuring larger fields.

Fig. 4 shows results of nuclear Rabi- as well as Larmor precession experiments at different driving fields. As nuclear spin $^{13}$C in the first shell of neighbors of the NV center are used. We have chosen these spins because they show the strongest interaction with the NV electron spin[30]. This facilitates the experiments since remanent fields from the write structure can randomly shift resonances by tens of MHz[26] which impairs addressing of more weakly coupled nuclear spin. The nuclear $^{13}$C spin is initialized via the electron spin by a frequency selective microwave $\pi$ pulse on one of the hyperfine transitions. As in the case of the electron spin after initialization a magnetic field at the frequency of the nuclear spin transition is applied. This frequency is determined by the strong hyperfine interaction of those nuclei to be 126 MHz. Fig. 4a shows Rabi oscillations of a single $^{13}$C with a frequency of 31.8 MHz. In analogy to the electron spin case, we also measured the $^{13}$C Larmor precession at different levels of B. We could achieve precession frequency changes of up to 14 MHz.

To reconstruct the magnetic field B from these values the impact of hyperfine coupling needs to be further analysed. Typically, nuclear spins have gyromagnetic ratios three orders of magnitude smaller than an electron spin. Specifically a free $^{13}$C has a $\gamma$ = 10.7 MHz/T and one would expect frequencies on the order of hundreds of kHz with the fields used in the experiments. However, the presence of the electron spin significantly alters the effective magnetic field at the location of the nuclear spin. This effective B field is given by the hyperfine interaction. Ideally, the interaction can be seen as an induced dipole that isotropically increases the magnetic field at the site of the nuclear spin such that the gyromagnetic ratio is $\gamma_{eff} = k_{hf}\gamma_n$, where the factor $k_{hf}$ depends on non-secular parts of the hyperfine interaction. However, first shell $^{13}$C spins have a highly anisotropic hyperfine interaction. We expand the NV Hamiltonian by a nuclear Zeeman term $H_{nB} = \gamma_n \vec{B}\vec{I}$ and a hyperfine term $H_{hf} = \vec{S}A\vec{I} = \sum_{i,j=x,y,z} A_{ij}S_iI_j$, where $A$ is the hyperfine tensor. With an anisotropic tensor the effective gyromagnetic ratio $\gamma_{eff} = \frac{1}{h}\frac{dE}{dB}$ becomes a function of the applied magnetic field strength and angle, where E are energies of the nuclear spin eigenstates. Thus, full knowledge of the hyperfine tensor is required for the evaluation of the magnetic field. We therefore measure the hyperfine tensor of the sensor spin in a separate experiment and use it to calculate the magnetic field B from the write head. From the Rabi measurements we derive a driving field amplitude of $B_{AC} = 47$ mT and the fastest precession corresponds to a field of $B_{DC} = 56$ mT tilted by 45° from the NV axis. The analysis is described in detail in the supplementary material[26].

Our results demonstrate that ultrafast coherent spin manipulations can be achieved with nanoscale magnetic structures over a bandwidth of almost 0.5 GHz. The large field gradients measured will be of use in quantum spintronic devices to e.g. locally drive electron spins in an array of interacting spins with distance on the order of few tens of nanometers[5]. Also nano MRI, using single NV centers as detectors will need ultrastrong field gradients which can be switched much faster than the Larmor precession of the nuclear spins to be investigated. The structures used in the current work are among those with the largest field gradients (10 mT/nm) and bandwidth available. Most importantly, single spins can be used to characterize such structures. While for current sizes simulations of their magnetic behaviour yield satisfactory results this is not the case once structure sizes approach 20nm. Hence, techniques are needed which are capable to characterize fields of such length scales in the presence of high background fields, large gradients and arbitrary orientation. Single defect centers appear to be an ideal tool as they are facile to operate and certainly meet the criteria for spatial resolution. With the present work, it is also evident that restrictions for NV magnetometry to measure large fields of arbitrary orientations can be overcome. Using multiple nuclear spins it is even possible to measure all vector components of a magnetic field simultaneously[26]. When entangling those spins nanoscale gradients can be determined with a single measurement. Combining cutting edge nano magnetism with single quantum probes thus seem to be an ideal match and profitable for future development of both techniques.

Acknowledgement

We acknowledge financial support by the German Science Foundation (SFB-TR 21, SPP1601 and FOR1493), the EU (SIQS), the DIADEMS consortium and the MPG. Furthermore, we thank Ilja Gerhardt, Amit Finkler and Helmut Fedder for their support and Jürgen Heidmann of Integral Solutions International for providing hard disk head samples and technical assistance.

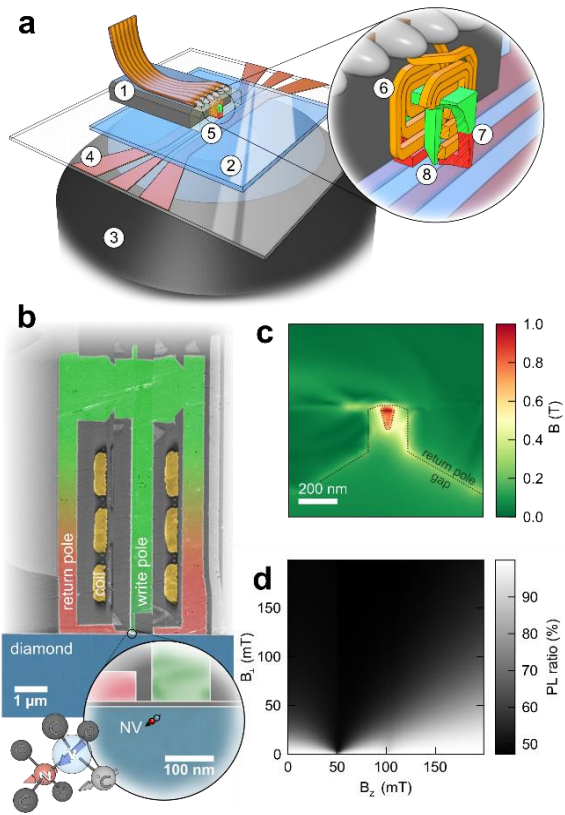

**Figure 1 Hard Disk Writer and NV**: **a** Schematics of the setup. A hard disk head (1) is in flat contact to the surface of a diamond membrane (2). An objective (3) granting optical access to NV centers and a microwave control structure (4) are located on the back side of the membrane. The writer (5) is embedded in the head. It consists of a set of coils (6) wrapped around high permeability return poles and central write pole. **b** Retouched cross-section SEM image of the writer. The central write pole is thinned down to tens of nanometers on its end. Return poles shape the magnetic field in the region of the pole to be highly local. The front return wraps around the write pole in three directions with a minimum distance of 20 nm. The head can be placed in contact with the diamond surface close to a shallow NV center. A model depicts the NV center with its neighboring carbon atoms including a paramagnetic $^{13}$C. **c** Finite element simulation of the magnetic field produced by the writer in a plane parallel to the surface in a 10 nm distance. The boundaries of the write pole and the front return pole are marked with dashed lines. **d** Numerical simulation of the NV photoluminescence depending on axial ($B_z$) and radial ($B_\perp$) magnetic fields. The simulation shows the steady state under continuous saturated pumping and is normalized to the zero-field photoluminescence.

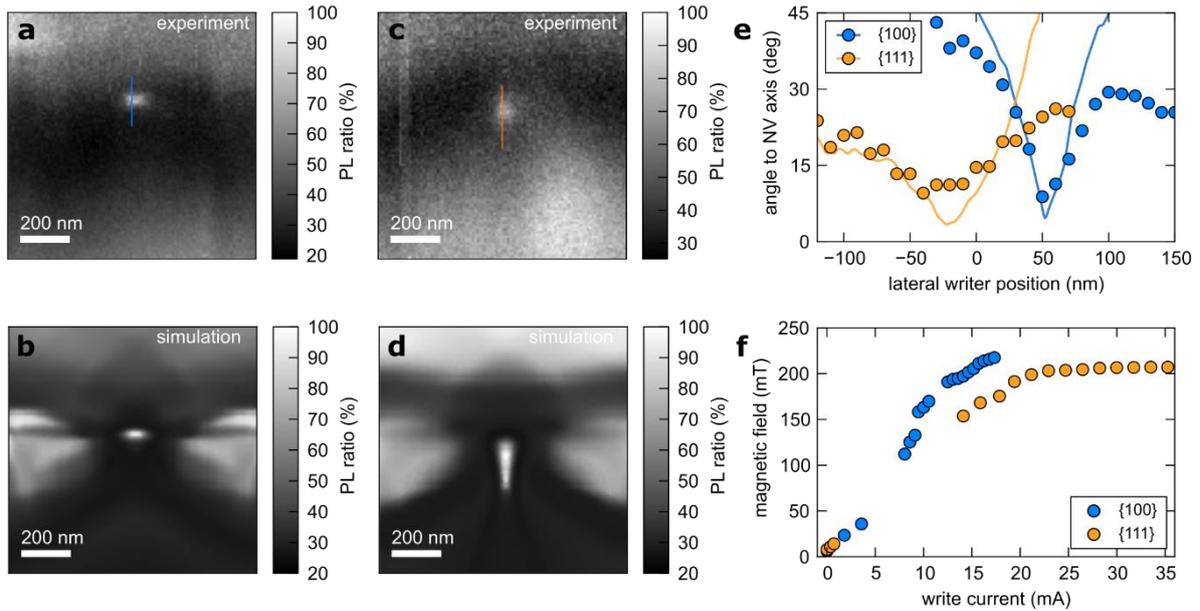

**Figure 2 Photoluminescence Imaging and Field Reconstruction**. **a** Measured gated PL of an NV center in a {100} sample depending on the relative position of the writer. The NV axis is tilted with 54.7° from the surface normal. In a spot of 50 nm size the magnetic field is almost aligned with the NV axis where it shows a significant increase in PL. **b** Simulation of the PL with the same geometry as in Fig. a. Apart from the alignment spot additional features appear on the sides, not reproduced in experiments. **c** Measured PL of an NV center in a {111} sample. The NV axis is parallel to the surface normal. The alignment spot appears straight underneath the pole. **d** Simulation for the same geometry. **e** Reconstruction of the field angle to the NV axis based on the PL of a line scan across the bright spots in a and c. The solid line shows the angles from the finite element simulation. The scatter points are reconstructed from the experimental PL data shown in Fig. a and c. **f** Magnetic field measured with ODMR on the bright spots for the two geometries ({100} blue, {111} orange) at different write currents. After a steep increase the field saturates at about 200 mT in both cases. While the {111} has a favorable lateral alignment position the {100} NV reaches larger values as it has a smaller axial distance to the pole (60 and 40 nm respectively).

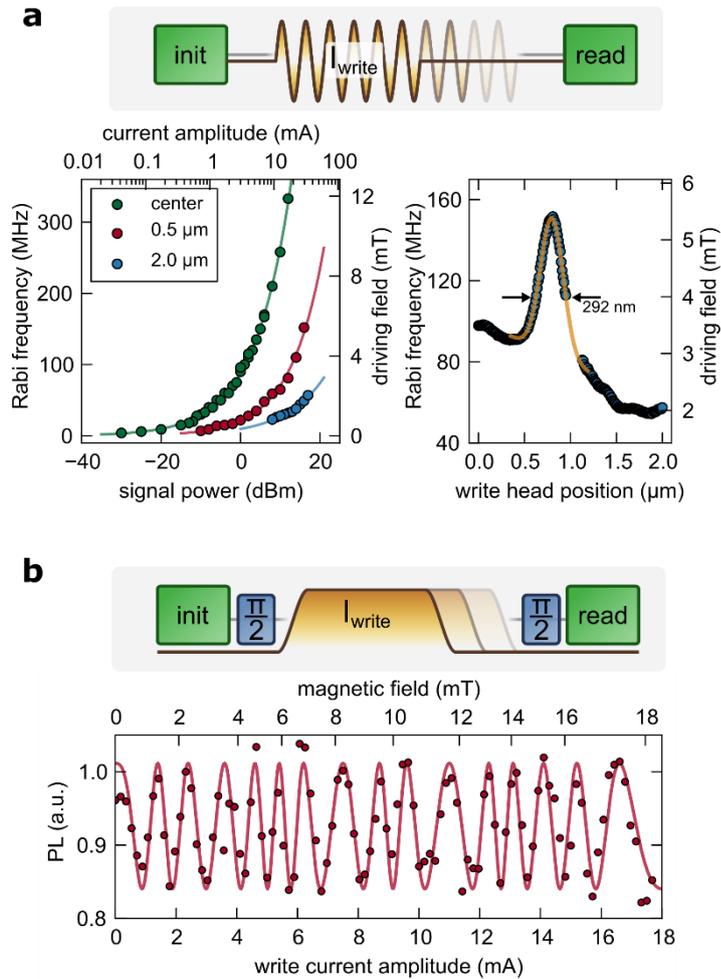

**Figure 3 Electron Spin Magnetometry**. **a** Applying a current to the write head modulated at 2.87 GHz, NV electron spin transitions can be driven. The spin is initialized and read out while the write current is off (pulse diagram top). The left panel shows Rabi frequencies and the derived magnetic fields over different signal powers for three different positions of the writer. With increasing signal power the Rabi frequencies increase up to a maximum of 333 MHz corresponding to a magnetic field amplitude of 6.2 mT perpendicular to the NV axis. The line plots are fits for a linear relation between the current amplitude and the magnetic field. The right panel shows a line scan of Rabi frequencies across the write pole with a fixed input power of 5 dBm with a 292 nm wide peak around the position of the write pole. The gradient reaches a maximum value of 18 µT/nm. **b** The writer is used to inject DC magnetic field pulses in the evolution period of a free induction decay experiment. A spin state superposition is prepared and the interference phase is read out while the write current is off (pulse diagram top). Sweeping through the current amplitude with fixed timing (evolution time 1500 ns, ramps 200 ns, DC plateau 800 ns) the increase in the phase evolution is recorded. An interpolation of the data (line plot) is used to derive a magnetic field scale. The data was recorded in a 500 nm lateral distance to the pole. All three experiments use {100} NVs with axis angles of 54.7° to the surface normal.

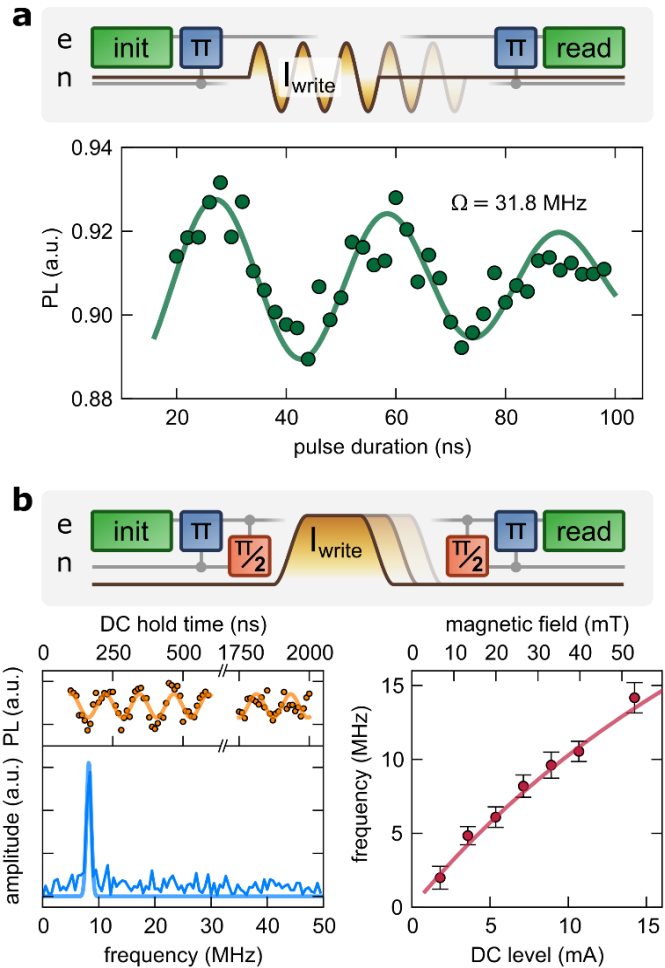

**Figure 4 Nuclear Spin Magnetometry**. **a** A single $^{13}$C spin can be driven with an AC field of the writer. The spin is prepared and read out while the current is off and a current is applied to the write head modulated at around 126 MHz is applied (pulse diagram top). A Rabi frequency of 31.8 MHz is achieved with a signal power of 20 dBm. **b** A free induction decay experiment with injected DC magnetic field pulses is conducted on the $^{13}$C nuclear spin. During the preparation of the spin superposition and the read out of the interference the writer is turned off (pulse diagram top). Using fixed evolution (3000 ns) and ramp (200 ns) times the DC plateau time is swept for different DC levels. The left panel shows an exemplary interference signal and its Fourier transform for a DC level of 7.1 mA. The right panel shows the phase evolution frequencies (dots) depending on the DC levels. The line plot shows simulated frequencies depending on the magnetic field. The magnetic field axis is linear and matched with ODMR measurements at low currents to the current axis. A {100} NV with an axis angle of 54.7° to the surface normal is used straight underneath the pole in both experiments.